# A combining earthquake forecasting model between deep learning and Epidemic-Type Aftershock Sequence(ETAS) model

Haoyuan Zhang[1], Shuya Ke[2], Wenqi Liu[1], Yongwen Zhang[1,*]
[1] Data Science Research Center, Faculty of Science, Kunming University of Science and Technology, Kunming, China.
E-mail: *zhangyongwen77@gmail.com*
[2] *School of Engineering, Hangzhou Normal University Hangzhou, China*

## Abstract

The scientific process of earthquake forecasting involves estimating the probability and intensity of earthquakes in a specific area within a certain timeframe, based on seismic activity laws and observational data. Epidemic-Type Aftershock Sequence (ETAS) models, which rely on seismic empirical laws, is one of the most commonly used methods for earthquake forecasting. However, it underestimates feature in short-term time scale and overestimates in long-term time scale. Convolutional Long Short-Term Memory (ConvLSTM), has emerged as a promising approach capable of extracting spatio-temporal features. Hence, we propose a novel composite model named CL-ETAS model, which combines the strengths of ConvLSTM and ETAS model. We conduct experimental verification on real seismic data in Southern California. Our results show that CL-ETAS model outperforms both ETAS model and ConvLstm in accurately forecasting the number of earthquake events, their magnitude, and spatial distribution. Additionally, CL-ETAS model demonstrates notable enhancements in forecast stability and interpretability. These results offer valuable insights and recommendations for future earthquake forecasting endeavors.

## 1. Introduction

Earthquakes are one of the most destructive natural disasters on the planet, posing a significant threat to life and property (Jordan *et al.* 2011). The scientific community has been actively studying earthquake forecasting, but the understanding in this area remains limited. Unlike other natural phenomena such as rainfall or hurricanes, which can be forecasted with some degree of accuracy, there is currently no unified mechanism for predicting when and where the next major earthquake will occur (Rikitake 1968). Different perspectives exist within the scientific community regarding earthquake forecasting. Some studies suggest that earthquakes cannot be reliably forecasted (Wyss 1997), while others propose various methods for forecasting future seismic events. These methods can be categorized based on the consideration of observable signals (Jordan 2006), statistical approaches (Gerstenberger *et al.* 2005), spatial tools (Jordan *et al.* 2011), classical machine learning algorithms and deep learning algorithms (DeVries *et al.* 2018), all operating under the assumption that earthquake events are forecastable.

There have been some attempts to forecast earthquakes using a diagnostic precursor via some observable signal, but so far, no successful and reliable forecast schemes have emerged (Jordan 2006; Jordan & Jones 2010; Ogata 2017). Currently, earthquake forecasts are primarily based on established seismic laws. The distribution of earthquake magnitudes follows the exponential Gutenberg-Richter law proposed by Gutenberg and Richter in 1944. According to the Utsu law, the number of earthquakes triggered by a mainshock increases exponentially with its magnitude (Utsu 1961). Additionally, the rate of triggered events decays over time following the power law known as the Omori law (Utsu & Ogata 1995). Empirical earthquake forecast schemes have been developed based on these laws, including the examination of earthquake clustering in space and time (Jordan & Jones 2010) and attributing spatiotemporal earthquake clusters to triggering events (Zaliapin *et al.* 2008). These schemes have led to the development of earthquake forecast models, such as the Epidemic-Type Aftershock Sequence (ETAS) model, which incorporates the Gutenberg-Richter, Utsu, and Omori laws into a Hawkes process. In this model, past earthquakes above a certain magnitude trigger subsequent earthquakes according to the same laws. Taroni et al. and Woessner et al. have demonstrated that clustering models, such as ETAS model, outperform other models in terms of forecast accuracy (Taroni *et al.* 2018; Woessner *et al.* 2011). Despite their contributions, these models have limitations in explaining certain fundamental characteristics of earthquakes (Vere-Jones 2005; Lippiello *et al.* 2008; Gulia *et al.* 2018; Gulia & Wiemer 2019). Furthermore, there are still unexplored statistical physics aspects related to seismic activity (de Arcangelis et al. 2016). To address these issues, some innovative approaches are being considered. Recent advancements by Zhang et al. (Zhang *et al.* 2020, 2021) have improved ETAS model by considering long-term memory behaviors of seismic activity.

The development of artificial intelligence and big data has garnered extensive attention across various industries, with machine learning being one of the fastest-growing technologies worldwide (Jordan & Mitchell 2015). In geological and earth science research, earthquake forecasting is one of the most challenging problems. Since the 1990s, machine learning has been utilized for earthquake forecasting. In recent years, many successful deep learning models have made significant breakthroughs in earthquake forecasting, mainly using Convolutional Neural Networks (CNN) and Long Short-Term Memory networks (LSTM). Kriegerowski et al. employed CNN as the primary model to locate 908 events by utilizing full waveform records of three components from multiple seismic stations (Kriegerowski *et al.* 2019). Li et al. introduced a deep learning model called DLEP for earthquake forecasting, which is based on eight precursor patterns and employs CNN with indicative and explicit features (Li *et al.* 2020). They also proposed a method for earthquake fault detection using CNN that requires only a small training set (Li *et al.* 2019). Huang et al. projected seismic events onto maps and generated an image dataset, labeling earthquakes with magnitudes of at least 7 as "1" and useing CNN to detect and forecast whether these major earthquakes would occur in the next 30 days (Huang *et al.* 2018). Wang et al. used LSTM to learn spatiotemporal correlations between earthquakes at different locations to make forecasts (Wang *et al.* 2017). Fabregas et al. developed a system based on various algorithms and LSTM that could forecast the frequency, maximum magnitude, and average depth of earthquake events in a specific area for a given year (Fabregas *et al.* 2020). Nicolis *et al.* (2021) used CNN to forecast the region where the largest earthquake event would occur and employed LSTM for forecasting the average number of earthquake events in the coming days based on the output of the intensity function estimated by ETAS model. Based on studies of CNN and LSTM, Shi et al. proposed the Convolutional Long Short-Term Memory (ConvLSTM) by combining both architectures (Shi *et al.* 2015). Fuentes and Alex González (Fuentes *et al.* 2022) introduced the Multi-column Convolutional Long Short-Term Memory (MConvLSTM) while studying the Chilean earthquake, which considers past crustal velocity values. They demonstrated that ConvLSTM is more accurate than other deep learning models for earthquake forecast.

At present, most earthquake forecast models for spatiotemporal seismic sequences rely on either empirical clustering models or deep learning models for spatiotemporal sequence forecasting. Some studies (Nicolis *et al.* 2021; Plaza *et al.* 2019) considered a deep learning model trained by the intensity function of ETAS model. Yet, they ignored the real seismic records to train the model. Hence, we provide a novel model for forecasting earthquakes named CL-ETAS model, which combines ETAS model, a well-established statistical model with empirical laws, with ConvLSTM, a deep learning model capable of processing spatiotemporal seismic records. This study aims to elucidate the individual strengths of these models and integrate them for improved spatiotemporal earthquake sequence forecasting.

## 2. Materials and methods

### 2.1. Data

The used dataset is the Southern California earthquake catalog, covering the time period from 1981 to 2022 which can be downloaded from SCEDC (Southern California Earthquake Data Center) at  https://scedc.caltech.edu/data/alt-2011-dd-hauksson-yang-shearer.html  (Hauksson *et al.* 2012).

The unit of magnitude used in this article is moment magnitude scale ($M_w$). For this catalog, earthquakes with magnitudes ($M \geq 3$) are complete, amounting to a total of 15,534 events. Below the magnitude ($M = 3$), the earthquake could be missing due to the weak detection of the seismic network.

### 2.2. ETAS model

Ogata proposed the ETAS model (Ogata 1988, 1998), which assumes that seismic events are a space-time stochastic point process in which each aftershock can potentially trigger further aftershocks. Each event above a magnitude $M_0$ is selected independently from the Gutenberg–Richter distribution (where $b = 1$). The conditional rate, $\lambda$, at location $(x, y)$ at time $t$ is given by

$$\lambda(x, y, t | H_t) = \mu(x, y) + \sum_{t_i < t} k(M_i) g(t - t_i) f(x - x_i, y - y_i, M_i),$$ *(1)*

where $H_t$ is the history process prior to $t$, $t_i$ are the times of the past events, and $M_i$ are their magnitudes $\mu(x, y) = \mu_0 u(x, y)$ is the background intensity at location $(x, y)$, where $u$ is the spatial probability density function (PDF) of background events, which is estimated by the method proposed by Zhuang (Zhuang *et al.* 2011; Zhuang 2012); $\mu_0$ is the background rate of the entire region. We represent the total number of the past events as $n - 1$.

The dependence of the triggering ability on magnitude is given by the Utsu law as

$$k(M_i) = A \ \exp(\alpha(M_i \text{-} M_0)),$$ *(2)*

where $A$ is the occurrence rate of earthquakes at zero lag. $\alpha$ is the productivity parameter.

The function $g(t - t_i)$ follows the Omori law as

$$g(t - t_i) = (1 + \frac{t - t_i}{c})^{-p},$$ *(3)*

where $c$ and $p$ are the Omori law parameters. The spatial clustering of aftershocks is implemented by introducing a spatial kernel function $f(x - x_i, y - y_i, M_i)$ (Zhuang 2012) as

$$f(x - x_i, y - y_i, M_i) = \frac{q - 1}{\pi \zeta^2}(1 + \frac{(x - x_i)^2 + (y - y_i)^2}{\zeta^2})^{-q},$$ *(4)*

where $\zeta = D \exp[\gamma_m(M_i - M_0)]$ indicates that the distances between triggering and triggered events depend on the magnitudes of the triggering events. $q$, $D$ and $\gamma_m$ are the estimated parameters.

The parameters fitted for ETAS model based on the R package "ETAS" (https://cran.r-project.org/package=ETAS) using the stochastic declustering approach introduced by Zhuang, Ogata, and Vere-Jones (Zhang *et al.* 2002), are as follows: $A = 10.73047$, $\alpha = 1.124465$, $M_0 = 3$, $\mu = 0.2554078$, $P = 1.065735$, $c = 0.003961263$. The spatial parameters were chosen to be $q = 1.378174$, $D = 0.001658387$ (in units of 'degree'), and $\gamma_m = 0.8093304$.

### 2.3. ConvLSTM

LSTM is particularly effective at forecasting significant occurrences with very long gaps and delays. However, for spatiotemporal sequence problems with strong spatiotemporal correlation, LSTMs do not perform well (Sutskever *et al.* 2014). While LSTMs excel at capturing time features, they are not as adept at capturing spatial features and cannot reflect spatial state information. To address this issue, Shi et al. proposed the ConvLSTM based on the principles of LSTMs (Shi *et al.* 2015). The ConvLSTM equation can be expressed as follows:

$$i_t = \sigma(W_{xi} * X_t + W_{hi} * H_{t-1} + W_{ci} \circ C_{t-1} + b_i),$$

$$f_t = \sigma(W_{xf} * X_t + W_{hf} * H_{t-1} + W_{cf} \circ C_{t-1} + b_f),$$

$$C_t = f_t \circ C_{t-1} + i_t \circ \tanh(W_{xc} * X_t + W_{hc} * H_{t-1} + b_c),$$

$$o_t = \sigma(W_{xo} * X_t + W_{ho} * H_{t-1} + W_{co} \circ C_t + b_o),$$

$$H_t = o_t \tanh(C_t), \qquad\qquad (5)$$

where $i_t$, $f_t$ and $o_t$ are the input gate, forget gate, and output gate, $C_t$ is the cell state, $H_t$ is the hidden state, $W$ are weights, $b$ are biases, $X_t$ is the number of input features, tanh is an activation function (hyperbolic tangent), $*$ denotes convolution operation and $\circ$ denotes Hadamard product.

In comparison to the LSTM equation, ConvLSTM modifies the input-to-state and state-to-state transitions by replacing them with convolution operations. This modification enables consideration of complex spatial structures. Additionally, when using convolution, larger kernels capture faster movements, while smaller kernels capture slower movements.

## 2.4. CL-ETAS model

Here, we present the structure and training process of the CL-ETAS model and compare it to the conventional ConvLSTM model.

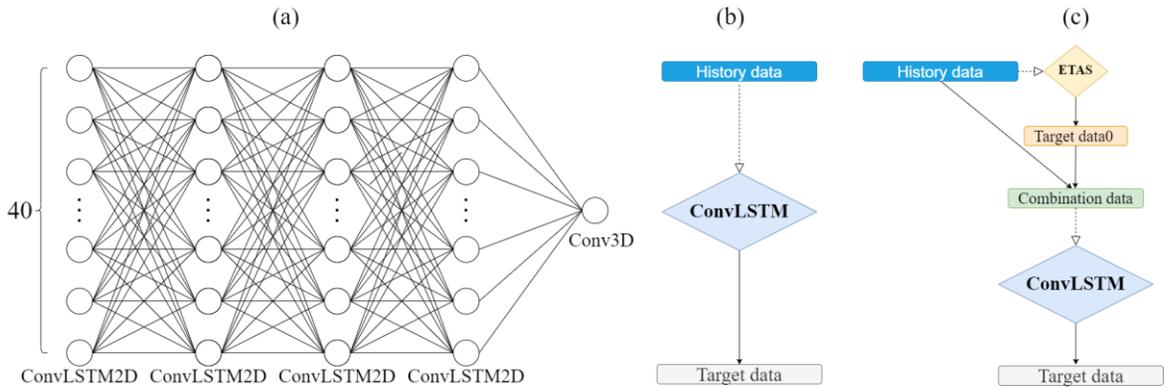

Figure 1: Schematics of (a) the neural network of ConvLSTM, (b) forecasting process of the conventional ConvLSTM model, and (c) forecasting process of the proposed CL-ETAS model.

Fig.1a illustrates the model structure of the neural network employed in this study for processing spatiotemporal sequential data, such as images or videos. The network is a ConvLSTM architecture consisting of four ConvLSTM2D layers and can be utilized for both forecasting and classification tasks. Each ConvLSTM2D layer incorporates 40 filters with a size of 3×3. The first layer features a specified input shape, and a Conv3D layer is added to the final layer to ensure that the output shape corresponds to the input shape. We deliberately chose a relatively simple network structure for this experiment to demonstrate that models trained using the proposed training process described in this paper exhibit improved generalization performance.

Fig.1b depicts the forecasting process of the conventional ConvLSTM, which involves feeding historical data directly into the model for training. The ConvLSTM employed in this study requires input data in the form of matrices. Accordingly, the earthquake catalog necessitates preprocessing and conversion into matrix format. As part of the training set, earthquake catalog data from 1981 to 1999 was selected. Relevant attributes such as magnitude, longitude, latitude, and time were extracted from the original earthquake catalog data, and these data were combined to form individual matrices. Specifically, the rows of the matrix represent grid index of earthquakes, i.e., we divide the entire region with a degree interval of two degrees for longitude and latitude into 20 grids and index them. The columns of the matrix with 5 columns represent magnitude index, with the first four starting from $M = 3$ to $M = 5$ and with a magnitude interval of $M = 0.5$, with the last one indicating $M \geq 5$ (as there are fewer earthquakes with $M = 5$, they are grouped into a single category). So, each element in the matrix for a specific day represents the number of earthquakes that occurred within a specific spatial region, the specific day and magnitude range. Then, the training set is consisted of 3263 matrices in chronological order from 1981 to 1999, each matrix has a size of 5 rows and 20 columns. Subsequently, a trained ConvLSTM was used for forecast. Since the output of magnitude for ConvLSTM is a rough range with the interval 0.5, we randomly generate the exact magnitude based on the Gutenberg-Richter distribution in this range.

Fig.1c presents the forecasting process of the proposed CL-ETAS model. Firstly, we employ the ETAS model to forecast target day. Subsequently, our approach involves the integration of target data0 derived from  ETAS model with historical data, leading to the creation of a combination dataset. To process this combination dataset in preparation for using ConvLSTM, we adopt the spatiotemporal matrix data preprocessing method as mentioned above. This methodology enables ConvLSTM to acquire knowledge not only from the intrinsic characteristics of historical data but also from the distinctive attributes of ETAS model, thereby facilitating an amalgamation of the two domains. Specifically, when training a model aimed at forecasting earthquakes one day in advance, the initial step entails employing CL-ETAS model to predict earthquakes occurring on the subsequent day. This iterative procedure is repeated numerous times, yielding multi-day forecasts generated by the model. Moreover, we can obtain many realizations of forecast target data based on different generations of ETAS model.

### 2.5. Forecast test

To evaluate the proposed model, we employ pycsep, a Python software package developed by the Collaboratory for the Study of Earthquake Predictability (CSEP) for earthquake prediction and data analysis (Savran *et al.* 2022). The package includes four Catalog-based Forecast Evaluations:

Number Test (N-Test), Magnitude Test (M-Test), Spatial Test (S-Test), and Pseudo-likehood Test (PL-Test). A brief description of each test and its purpose is provided below.

### 2.5.1. Catalog-based forecast tests

Define a region $R$ as a function of some magnitude range $M$, spatial domain $S$ and time period $T$

$$R = M \times S \times T .$$  (6)

An earthquake can be described by a magnitude $m_i$ at some location $s_j$ and time $t_k$. A catalog is a collection of earthquakes, thus the observed catalog can be written as

$$\Omega = \{e_n | n = 1 \ldots N_{obs}; e \in R\},$$  (7)

and a forecast is then specified as a collection of synthetic catalogs containing events $\hat{e}_{nj}$ in domain $R$, as

$$\Lambda \equiv \Lambda_j = \{\hat{e}_{nj} | n = 1 \ldots N_j, j = 1 \ldots J; \hat{e}_{nj} \in R\}.$$  (8)

A forecast consists of $J$ simulated catalogs each containing $N_j$ events, described in time, space and magnitude such that $\hat{e}_{nj}$ describes the $nth$ synthetic event in the $jth$ synthetic catalog $\Lambda_j$.

### 2.5.2. N-tests

The purpose of the N-Test is to assess whether the forecast aligns with the number of observed events.

In this case, the observed statistic is denoted by $N_{obs} = |\Omega|$ and represents the number of events recorded in the observed catalog.

$$N_j = |\Lambda_j|; j = 1 \ldots J .$$ (9)

The probabilities of $N$ events or less and $N$ events or more can be evaluated using the empirical cumulative distribution function of $F_n$:

$$\alpha_1 = P(N_j \geq N_{obs}) = 1 - F_N(N_{obs} - 1) ,$$

$$\alpha_2 = P(N_j \leq N_{obs}) = F_N(N_{obs}) .$$ (10)

### 2.5.3. M-tests

The purpose of M-Test is to assess how well the observed frequency-magnitude distribution aligns with the simulated catalogs used for forecasting.

The union catalog $\Lambda_u$ can be established by combining all the simulated catalogs featured which can be expressed formally as follows:

$$A_U = \{\lambda_1 \cup \lambda_2 \cup \ldots \cup \lambda_j\} .$$ (11)

the union catalog contains all events across all simulated catalogs for a total of $N_u = \sum_{j=1}^{J} |\lambda_j|$.

The observed statistic can be computed:

$$d_{obs} = \sum_k (\log[\frac{N_{obs}}{N_U} \Lambda_U^{(m)}(k) + 1] - \log[\Omega^{(m)}(k) + 1])^2 ,$$ (12)

where the histogram of the union catalog magnitudes $\Lambda_U^{(m)}(k)$ and the histogram of the union catalog magnitudes $\Omega^{(m)}(k)$ represent the count in the *kth* bin of the magnitude-frequency distribution.

### 2.5.4. S-tests

The purpose of S-Test is to evaluate the agreement between the spatial rates in the forecast and the observed events by isolating the spatial component of the forecast.

In this case, we use the normalisation $\hat{\bar{\lambda}}_s = \hat{\lambda}_s / \sum_R \hat{\lambda}_s$. Then the observed spatial test statistic is calculated as

$$S_{obs} = [\sum_{i=1}^{N_{obs}} \log \hat{\bar{\lambda}}_s^*(k_i)]N_{obs}^{-1},$$    *(13)*

in which $\hat{\bar{\lambda}}_s^*(k_i)$ is the normalised approximate rate density in the *kth* cell corresponding to the *ith* event in the observed catalog $\Omega$.

### 2.5.5. PL-tests

The purpose of PL-Test is to evaluate the likelihood of a forecast given an observed catalog.

A continuous marked space-time point process can be specified by a conditional intensity function $\lambda(e|H_t)$, in which $H_t$ describes the history of the process in time. The log-likelihood function for any point process in $R$ is given by

$$L = \sum_{i=1}^{N} \log \lambda(e_i|H_t) - \int_R \lambda(e|H_t)dR.$$    *(14)*

We approximate the expectation of $\lambda(e|H_t)$ using the forecast catalogs. The approximate rate density is defined as the conditional expectation given a discretised region $R$ of the continuous rate

$$\hat{\lambda}(e|H_t) = E[\lambda(e|H_t)|R_d].$$ *(15)*

The pseudo-loglikelihood is then

$$\hat{L} = \sum_{i=1}^{N} \log \hat{\lambda}(e_i|H_t) - \int_R \hat{\lambda}(e|H_t) dR.$$ *(16)*

and we can write the approximate rate density as

$$\hat{\lambda}(e|H_t) = \sum_M \hat{\lambda}(e|H_t),$$ *(17)*

where we take the sum over all magnitude bins $M$. We can calculate observed pseudo likelihood as

$$\hat{L}_{obs} = \sum_{i=1}^{N_{obs}} \log \hat{\lambda}_s(k_i) - \bar{N},$$ *(18)*

where $\hat{\lambda}_s(k_i)$ is the approximate rate density in the $kth$ spatial cell and $k_i$ denotes the spatil cell in which the $ith$ event occurs. $\bar{N}$ is the expected number of events in $R_d$.

## 3. Results

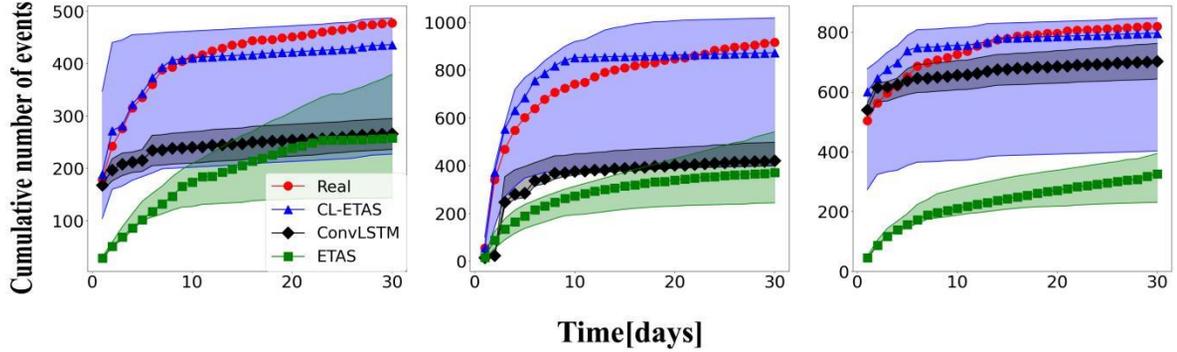

Figure 2: Forecasting of cumulative number of aftershocks ($M \geq 3$) as a function of time for (a), the 1999 Hector Mine earthquake ($M = 7.1$) as E1, (b) the 2010 Baja California earthquake ($M = 7.2$) as E2 and (c), the 2019 Ridgecrest earthquake ($M = 7.1$) in the Southern California. The shaded areas correspond to the 95% confidence intervals of 500 forecasting realizations.

In this study, we use the training process to obtain the forecasting model as CL-ETAS model. We compared the performance of CL-ETAS model, ConvLSTM, and ETAS model in forecasting the number of earthquake events after large shocks. For Southern California, a large earthquake ($M \geq 7$) occurs approximately every ten years. Here, we select three mainshocks with $M \geq 7$ that occurred in the testing dataset between 1999 and 2019 as examples. These included the 1999 Hector Mine earthquake (referred to as E1, $M = 7.1$), the 2010 Baja California earthquake (referred to as E2, $M = 7.2$), and the 2019 Ridgecrest earthquake (referred to as E3, $M = 7.1$). The forecasting results are illustrated in Fig. 2.

Our results show that the ETAS model exhibits a gradually increasing trend but overall underestimates the number of earthquake events in comparison to that of real data in Fig. 2. Especially, Fig. 2a shows that the forecasting results of ETAS model approached the real data gradually in long time (above 20 days). It implies that ETAS model has the underestimated feature in short-term time scale but overestimates in long-term time scale. This could be due to that ETAS model has no some physical features such as the varied productive law with time (Zhang *et al.* 2020, 2021). In Fig. 2b and 2c, ETAS model also significantly underestimate the number of earthquake events.

On the other hand,  ConvLSTM displays an overall trend of first rising and then slowly increasing, with  better performance than ETAS model. However, ConvLSTM still underestimates the number of earthquake events, especially from 3 to 10 days in Fig. 2.  The forecasting results of ConvLSTM are close to and even above the real values in the early days as shown in Fig. 2c. We conjecture that ConvLSTM indeed learns some earthquake features. But duo to some limitations of training process (i.e., insufficient dataset), ConvLSTM does not learn well about the time evolution laws of earthquakes such as the Omori law.

To outcome the deficiencies of ETAS model and ConvLSTM, we thus introduce CL-ETAS model, which incorporates features of both ETAS model and ConvLSTM. Our results show that CL-ETAS model have the trend consistent with the trend of real data as shown in Fig. 2. The real numbers are within the confidence intervals of the forecasting of CL-ETAS, which is bigger than those of ETAS and ConvLSTM, since the uncertainty of forecasting of CL-ETAS model depends on both ETAS and ConvLSTM. Since CL-ETAS model includes ETAS model forecasting data to forecast such that captures the empirical laws in ETAS model, CL-ETAS model exhibits a better performance in comparison to both the forecasting results of  ETAS model and ConvLSTM.

Subsequently, we conduct a comparison of the CL-ETAS model, ConvLSTM, and ETAS model in forecasting the spatial distribution of earthquakes over time. The results are presented in Figs. 3-5 for different forecasting time windows. Fig. 3 demonstrates that CL-ETAS model offers a better forecast performance compared to other two models. This outcome is mainly attributed to the significant advantages of CL-ETAS model in earthquake count forecasting. Furthermore, CL-ETAS model exhibits an ability to accurately forecast high-magnitude earthquakes, which is not achievable by ConvLSTM and ETAS models.

In Figs. 4 and 5, both ConvLSTM and ETAS models clearly demonstrate their respective advantages as the time window increases. ConvLSTM outperforms ETAS model in forecasting earthquake magnitudes and occurrence counts, as depicted in Fig.4h, 4k, Fig.5h, and 5k. The earthquake magnitudes and occurrence counts forecasted by ConvLSTM are closer to the real data, while ETAS model fails to forecast larger magnitudes and exhibits a noticeable deficiency in counts. However, in terms of forecasting the spatial distribution of earthquakes, ConvLSTM falls short compared to the performance of ETAS model, as shown in Fig.4i, 4l, Fig.5g, and 5j. ConvLSTM's forecasts tend to be biased towards the epicenter, neglecting aftershocks that occur at locations far away from the main earthquake. On the other hand, ETAS' forecasts appear to be more accurate and evenly distributed. In contrast, CL-ETAS model combines the strengths of both models and demonstrates outstanding performance characteristics. The results indicate that CL-ETAS model performs exceptionally well in forecasting earthquake counts and magnitudes, while also achieving a spatial distribution closer to the real data. However, during the training of the model, we construct the input matrices with two degrees' resolution such that the spatial distribution of forecasting earthquakes for CL-ETAS and ConvLSTM model is coarse.

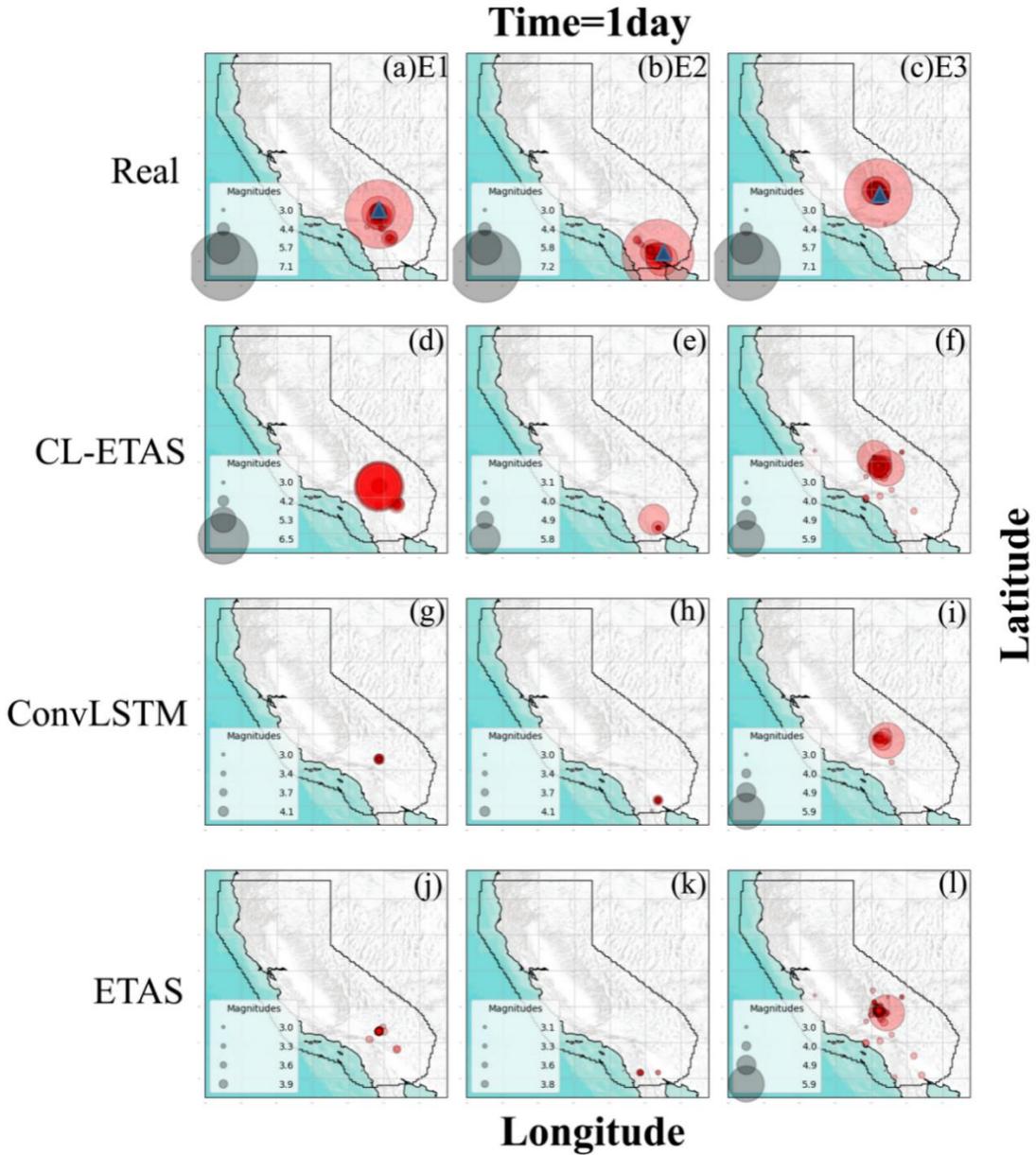

Figure 3: Forecasting of spatial distribution of aftershocks within 1 day after 3 mainshocks E1, E2 and E3 for (a)-(c) real (including mainshocks as blue triangle with largest circle), (d)-(f) CL-ETAS, (g)-(i) ConvLSTM, and (j)-(l) ETAS . The size of the circle denotes the magnitude of the earthquake. A larger circle indicates a higher magnitude.

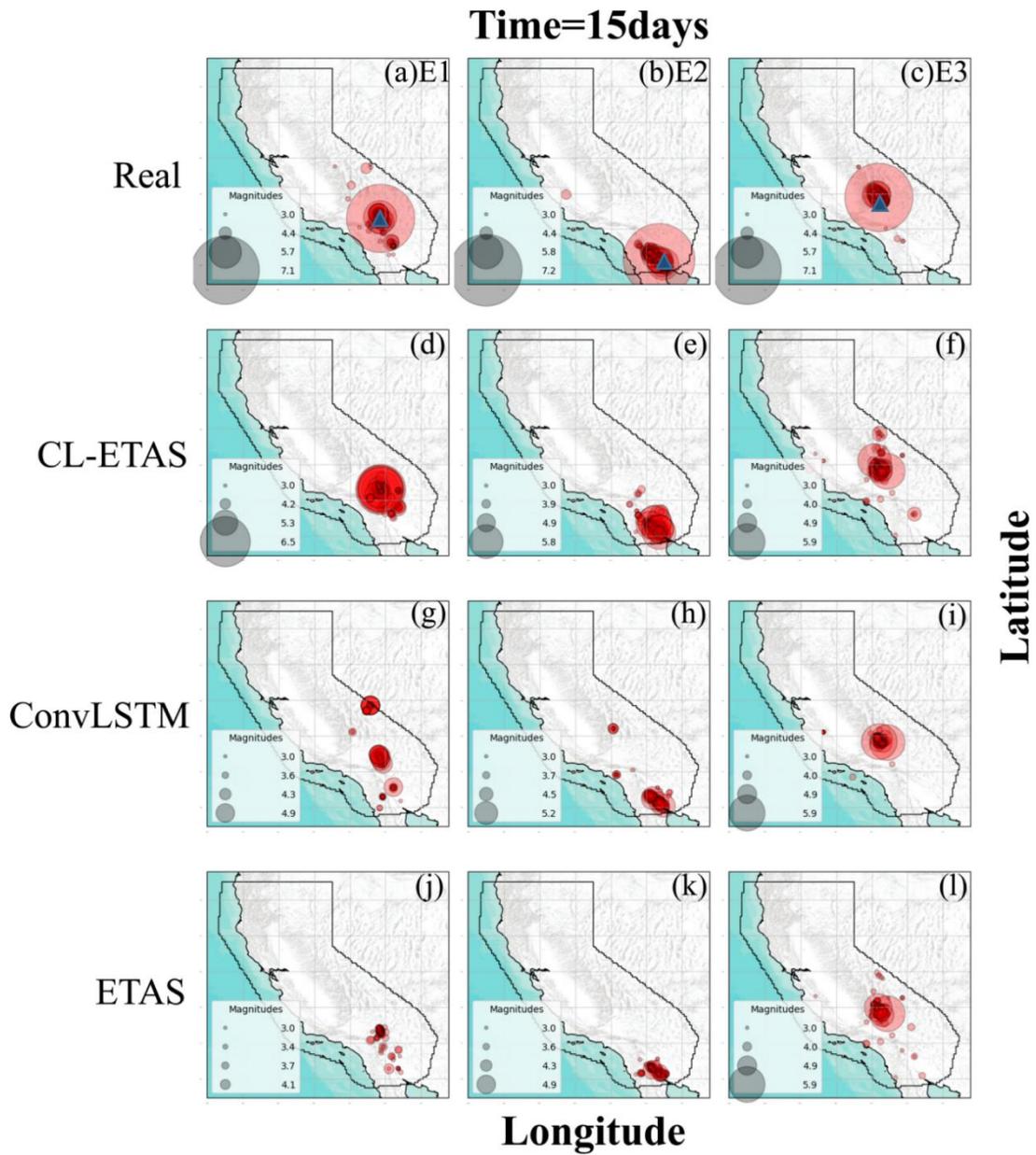

Figure 4: Same as Figure 3 but for 15 days

**Time=30days**

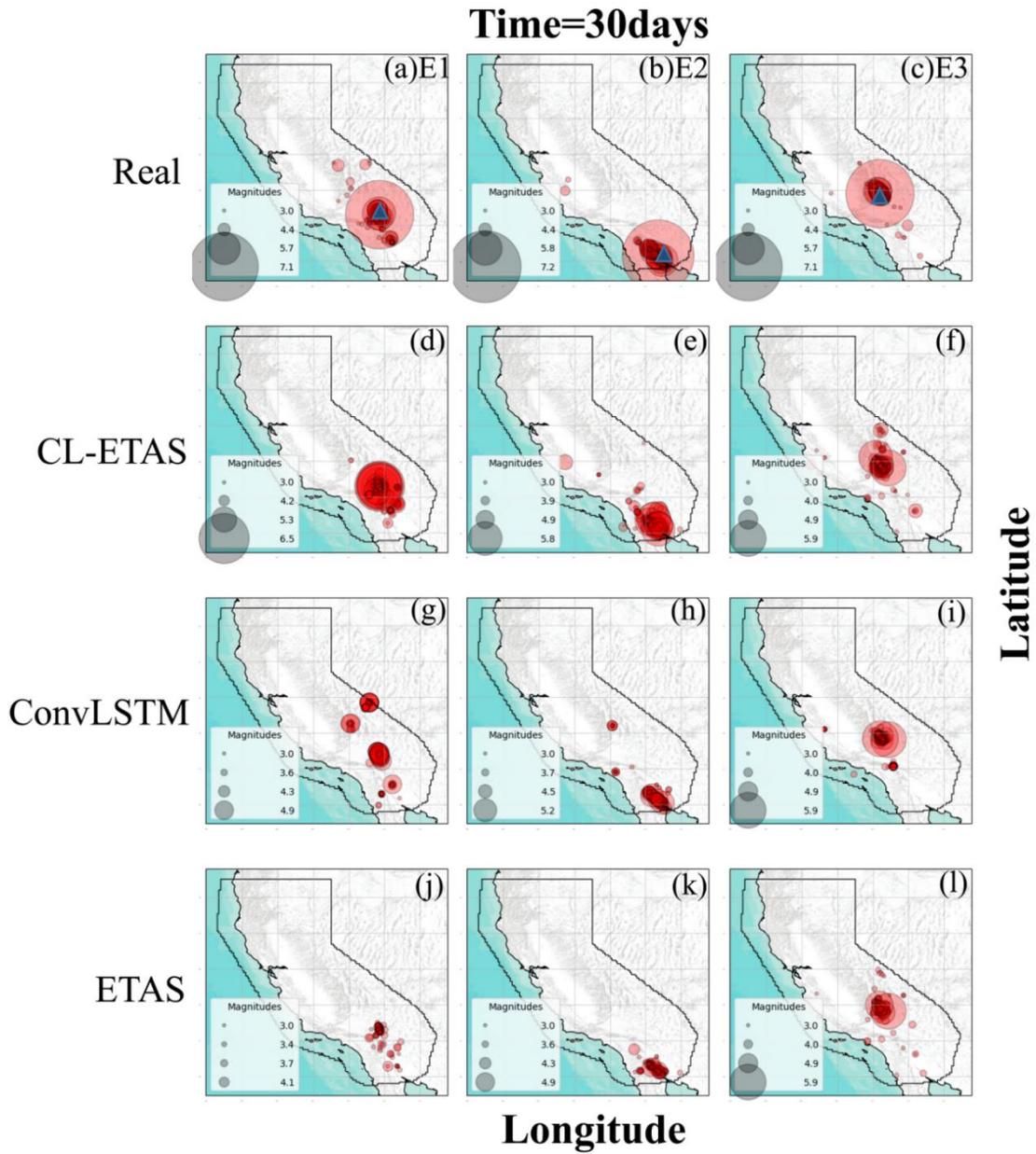

Figure 5: Same as Figure 3 but for 30 days.

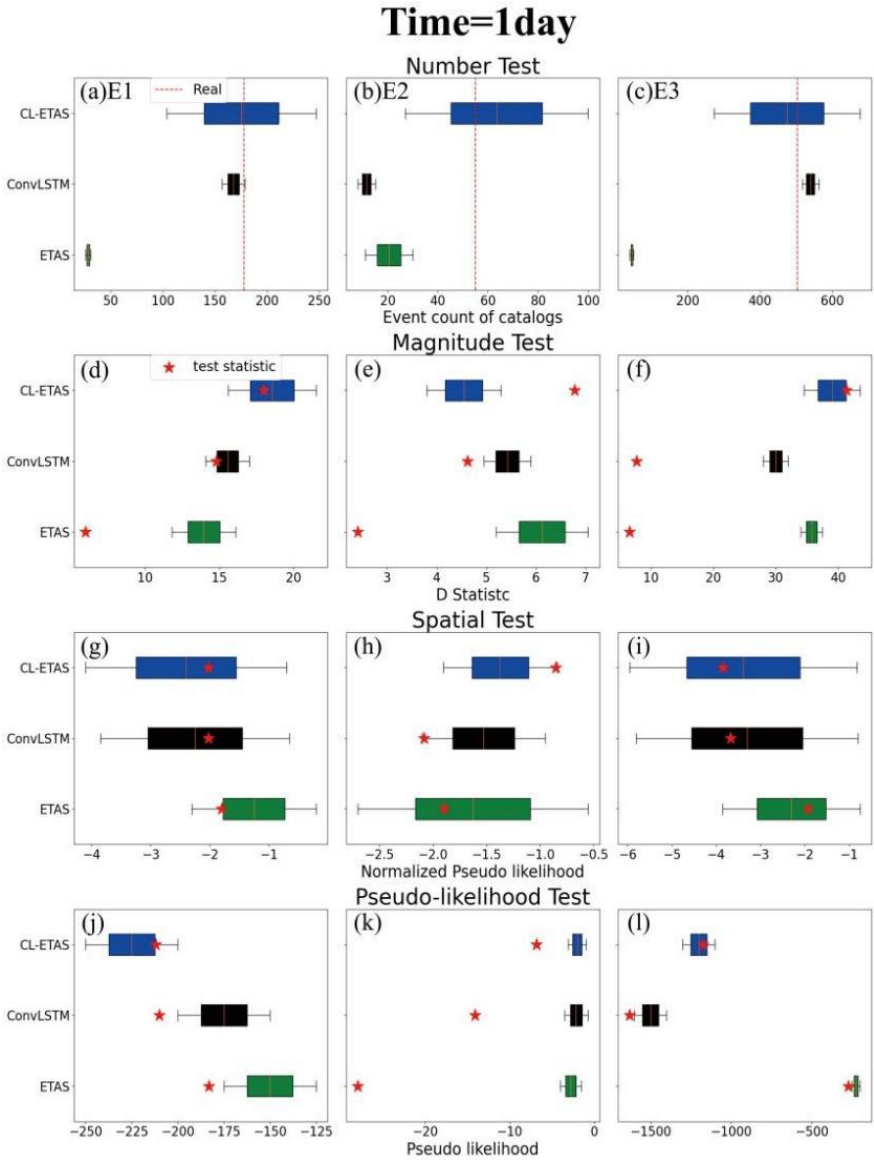

Figure 6: Boxplot of forecasting model evaluation for aftershocks of E1, E2 and E3 within 1 day for (a)-(c) number test (the red dashed line represents the real number), (d)-(f) magnitude test, (g)-(i) spatial test, and (j)-(l) pseudo-likelihood test. The red star indicates the test statistic to evaluate the performance of an earthquake forecast model. When the red star is closer to the median, the performance is better.

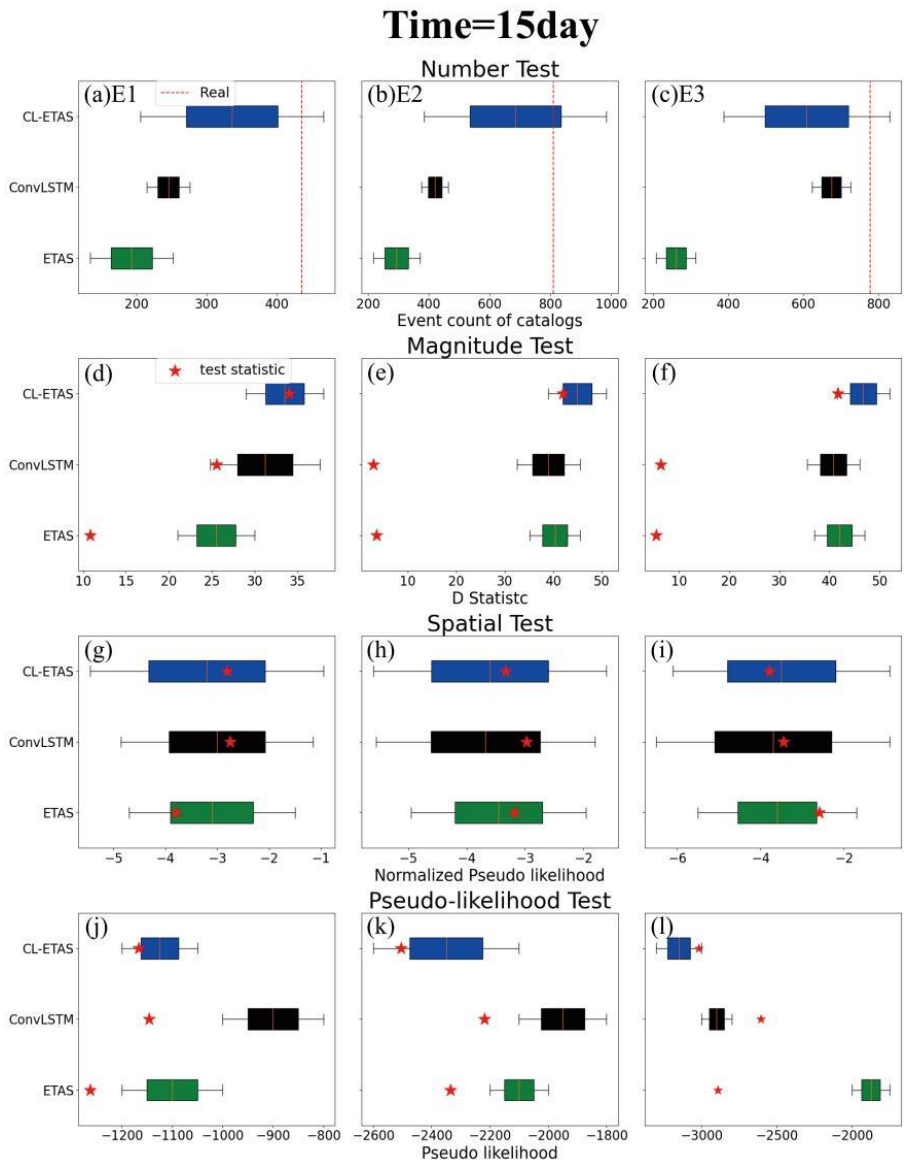

Figure 7: Same as Figure 6 but for 15 days.

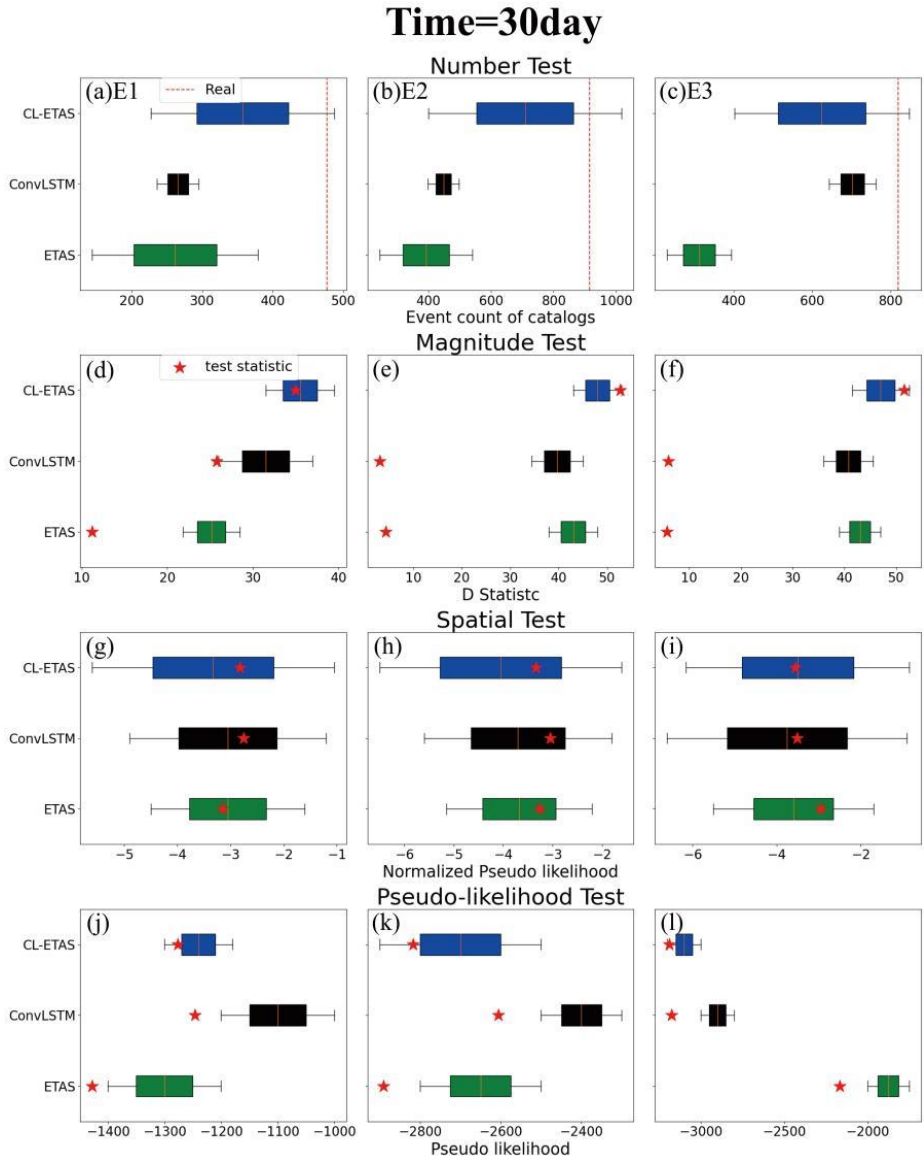

Figure 8: Same as Figure 6 but for 30 days.

Next, we compare the performance of the three models in forecasting earthquake occurrences after three mainshocks over three time periods. To evaluate forecast accuracy, we employ four catalog forecasting evaluation criteria: N-Test, M-Test, S-Test, and PL-Test (see Materials and methods). The results are presented in Figs. 6-8 for different forecasting time windows.

The result of the N-Test in Figs. 6-8 indicates that CL-ETAS model passes all tests in the 1-day, 15-day, and 30-day forecasts. In contrast, ConvLSTM model only passes the test in the 1-day forecast in Fig. 6a, and ETAS model does not pass any of the tests in three time windows.

The result of the M-Test in Figs. 6-8 indicates that CL-ETAS model passes all tests in the 1-day, 15-day, and 30-day forecasts, with the exception of Fig. 6e. This discrepancy is due to the limited amount of true data available on that day, resulting in a corresponding decrease in forecastable results and a slight deviation in earthquake magnitude forecasts. However, this deviation remains small and falls within the normal range of error. In contrast, ConvLSTM passes the test in the 1-day forecast in Fig. 6d, the test in the 15-day forecast in Fig. 7d, and the test in the 30-day forecast in Fig. 8d. We find that although ConvLSTM passes three tests, it only barely manages to do so. This occurrs because the aftershocks triggers by E1 are relatively concentrated in magnitudes range ( $4.5 \geq M \geq 3.5$) within 30 days, and ConvLSTM happens to perform well in forecasting these magnitudes. ETAS model does not pass any of the tests in three time windows.

The results of the S-Test in Figs. 6-8 indicates that all models successfully passes the tests in the 1-day, 15-day, and 30-day forecasts. Moreover, the test statistics (see Materials and methods) shown as the red stars in Figs. 6-8 indicate that the performance of CL-ETAS model remains superior to the other two models.

The result of the PL-Test in Figs. 6-8 indicates that CL-ETAS model passes all tests in the 1-day, 15-day, and 30-day forecasts. In contrast, ConvLSTM and ETAS models only partially pass the test in the 1-day forecast (as shown in Fig. 6l).

In conclusion, our investigation has found that CL-ETAS model passes all tests for in the 1-day, 15-day, and 30-day forecasts, successfully demonstrating its rationality and correctness. Additionally, this model exhibits excellent performance in forecasting earthquake counts and spatial distribution. Our findings indicate that CL-ETAS model represents a remarkable improvement over ConvLSTM and ETAS models in terms of accuracy and forecastable performance.

## 4. Conclusions and discussions

In this paper, we introduce a novel earthquake forecast model known as CL-ETAS model, which combines the strengths of the deep learning model ConvLSTM and the statistical model ETAS model. Our results show that CL-ETAS model have the trend consistent with the trend of real data for aftershocks of three mainshocks (the 1999 Hector Mine earthquake ($M = 7.1$); the 2010 Baja California earthquake ($M = 7.2$); the 2019 Ridgecrest earthquake ($M = 7.1$)) in Southern California. The real numbers are within the confidence intervals of the forecasting of CL-ETAS model. CL-ETAS model exhibits a better performance in comparison to both the forecasting results of ETAS model and ConvLSTM. CL-ETAS model offers markedly superior forecast performance compared to the other two models in forecasting the spatial distribution and magnitude of earthquakes over time. Additionally, we test all three models using N-Test, M-Test, S-Test, and PL-Test in the 1-day, 15-day, and 30-day forecasts, and results indicate that CL-ETAS model passes all tests and performs the best overall.

To further enhance the performance of CL-ETAS model, we can consider improvements in two aspects. Firstly, concerning the usage of ConvLSTM, we can try increasing the number of neural network layers, selecting more complex structures, or modifying other parameters such as learning rate, batch size, and optimizer type. Secondly, regarding the usage of ETAS model, we can adjust the time window, spatial range, and the minimum or maximum magnitude of earthquake activity when generating target data with ETAS model, in order to create more accurate data.

The CL-ETAS model has been shown to be superior, effective, and reliable. It has also shown significant potential in terms of mitigating seismic dangers. The future research will focus on further exploring and applying this model to achieve better results and wider applications.

**ACKNOWLEDGMENTS**

The authors thank the data source provided by website (https://scedc.caltech.edu/data/alt-2011-dd-hauksson-yang-shearer.html), the National Natural Science Foundation of China (No. 12305044 and 12371460) and the Fundamental Research Program of Yunnan Province (No. CB22052C173A).

**CONFLICTS OF INTEREST**

The authors declare that they have no known competing financial interests or personal relationships that could have appeared to influence the work reported in this paper.

**DATA AVAILABILITY**

The data used in this study are all open-access. The data can be downloaded from SCEDC (Southern California Earthquake Data Center) at https://scedc.caltech.edu/data/alt-2011-dd-hauksson-yang-shearer.html.